\documentclass[a4paper, oneside, twocolumn, notitlepage, 10pt]{extarticle_ecoc2015}
\usepackage{ecoc2015}
\usepackage{amsmath,amssymb}
\usepackage{mathtools, cuted}
\usepackage{graphicx}
\usepackage{float}
\usepackage{pgfplots}
\usepackage{caption}
\usepackage{tikz}
\usepackage{pgfplots}
\usepackage{tikz}
\usetikzlibrary{plotmarks}
\usepackage{xcolor}
\usepackage{hyperref}
\usetikzlibrary{shapes}
\usepackage{pgfplotstable}
\usepackage{stfloats}
\usetikzlibrary{plotmarks}
\usetikzlibrary{arrows}
\usetikzlibrary{shapes,decorations}
\usetikzlibrary{colorbrewer}
\usepackage{caption}
\pgfplotsset{compat=1.14}
\usetikzlibrary{plotmarks}
\usetikzlibrary{external}
\usepgfplotslibrary{external} 
\begin{document}
\selectlanguage{american}    % Standard Language

%-------------------------------------------------- Title -----------------------------------------------------%

\title{The ISRS GN Model, an Efficient Tool in Modeling Ultra-Wideband Transmission in Point-to-Point and Network Scenarios}%

%------------------------------------------------- Authors-----------------------------------------------------%

\author{
Daniel Semrau\textsuperscript{(\textasteriskcentered)}, Eric Sillekens, Robert I. Killey, Polina Bayvel
}

\maketitle                  % Create title and author

%------------------------------------------ Description of Authors ----------------------------------------------%

\begin{strip}
 \begin{author_descr}

   Optical Networks Group, Department of Electronic and Electrical Engineering, University College London (UCL), Torrington Place, London WC1E 7JE, UK,
   \textsuperscript{(\textasteriskcentered)} \uline{uceedfs@ucl.ac.uk}
 \end{author_descr}
\end{strip}

\setstretch{1.1}

%-------------------------------------------------- Abstract ---------------------------------------------------------%

\begin{strip}
  \begin{ecoc_abstract}
An analytical model to estimate nonlinear performance in ultra-wideband optical transmission networks is presented. The model accurately accounts for inter-channel stimulated Raman scattering, variably loaded fibre spans and is validated through C+L band simulations for uniform and probabilistically shaped 64-QAM.
  \end{ecoc_abstract}
\end{strip}

%introduction-------------------------------------------------------%
\section{Introduction}
Accurate analytical models to estimate the overall system and network performance in the presence of nonlinear distortion are key for efficient system design and network optimization. In paticular, the conventional Gaussian Noise (GN) model has been extensively applied in point-to-point and network scenarios in order to derive optimal system operation or networking strategies \cite{GNmodel,Ives}. 
To increase overall network throughput, the potential of ultra-wideband transmission
is being increasingly explored.  Associated with bandwidth increases is the power transfer due to inter-channel stimulated Raman scattering (ISRS), which can become significant\cite{Semrau}.
New network design and estimation strategies are needed to take ISRS into account in the estimate of nonlinear interference (NLI) noise, used to approximate nonlinear distortion, especially in the cases of wavelength routed networks. In this case, wavelength-routing and lightpath assignment algorithms determine where lightpaths are added and dropped, and these are designed subject to  overall performance metric and network topology.
The impact of ISRS on the NLI has been experimentally assessed over continuous bandwidths of 3\cite{Cantono2018} and 9\cite{SaavedraSRS}~THz and between S- and L-band\cite{MinoguchiSRS} in point-to-point transmission, showing good agreement with theoretical predictions. To date, however, the case of variably loaded network spans, carrying different combinations of lightpaths/wavelegth channels has not been considered.
\ 
\par
In this work, an analytical model for NLI estimation is presented which accounts for ISRS and variably loaded network spans as a result of wavelength-routing. It is applied to a point-to-point and a network scenario and validated by split-step simulations over C+L band using uniform and probabilistically shaped QAM signals.
\begin{figure}[h]
\includegraphics[]{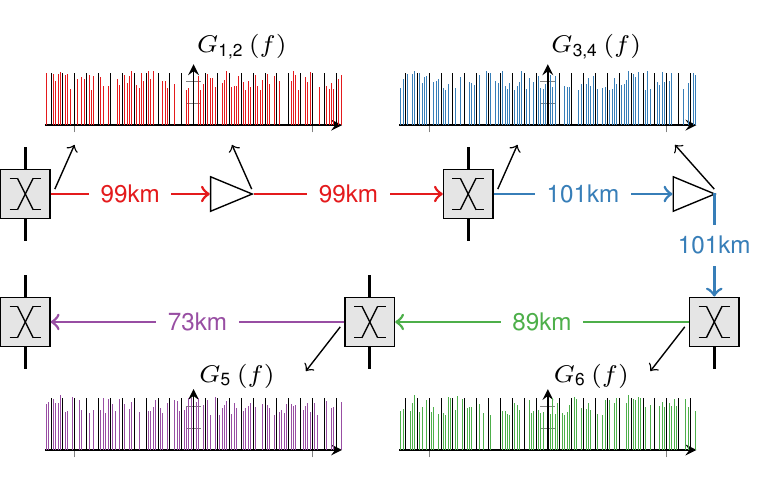}
\caption{Example transmission link, between nodes 15 and 13, from the BT~20+2 network topology\cite{Ives}, showing interfering channels (in colour) added and dropped at each ROADM.}
\label{fig:networkscheme}
\end{figure}
%

%methods -------------------------------------------------------%
\section{The ISRS GN model}
\begin{figure*}[h]
\centering
\includegraphics[]{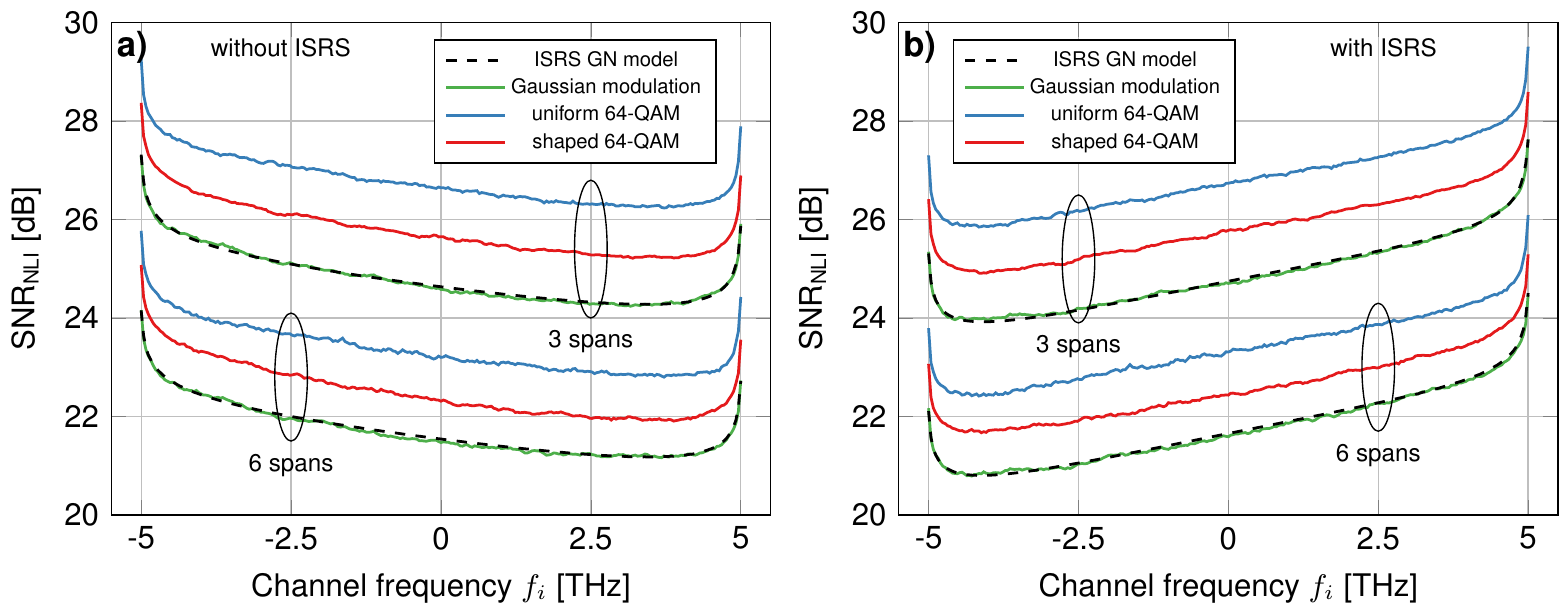}
\caption{\small Nonlinear performance of a fully loaded C+L band point-to-point scenario a) without and b) with ISRS.}
\label{fig:straight}
\end{figure*}
Neglecting linear noise contributions, the signal-to-noise ratio resulting from NLI (i.e. the nonlinear performance) is given by
\begin{equation}
\begin{split}
\text{SNR}_{\text{NLI},i} = \frac{P_i}{\sigma^2_{\text{NLI},i}} = \frac{P_i}{\int G\left(f+f_i\right)df},
\label{eq:SNR}
\end{split}
\end{equation}
with launch power $P_i$, carrier frequency $f_i$ of channel $i$ and NLI power spectral density (PSD) $G\left(f\right)$. The integration in \ref{eq:SNR} is carried out over the channel bandwidth.
\ 
\par
Recently, we presented a GN model which includes the effect of ISRS and is referred to as the ISRS GN model \cite{SemrauSRS}. Comparable results were reported in \cite{Semrau,Cantono2018}.
\begin{figure*}[b]
\vspace*{-0.5cm}
\begin{equation}
G(f) = \frac{16}{27}\gamma^2 \ G_1(f)\int df_1\int df_2\
\cdot\left|\sum_{k=1}^{n} \ \int_{0}^{L_k} d\zeta \ S_k\left(f_1,f_2,f\right) \frac{P_{k}e^{-\alpha \zeta-P_{k}C_{\text{r}} L_{\text{eff}}(f_1+f_2-f)}}{\int G_{k}(\nu)e^{-P_{k}C_{\text{r}} L_{\text{eff}}\nu} d\nu}e^{j\phi\left(f_1,f_2,f,\tilde{L}_k+\zeta\right)}\right|^2
\label{eq:ISRSGNmodel}
\end{equation}
\end{figure*}
\ 
\par 
Eq. (2) is an extension of the ISRS GN model to account for variably loaded fibre spans. A different signal PSD for each span in a point-to-point transmission can be the result of non-ideal gain equalization or gain-ripples of the optical amplifiers. In a network context, varying signal PSD's are additionally the result of a given network load that potentially leads to unused channel slots between reconfigurable optical add-drop multiplexers (ROADM).
\ 
\par 
The ISRS GN model only requires knowledge of the signal PSD $G_k\left(f\right)$ at the input of span $k$ which can be accessed through spectral power monitoring. The PSD's are then inserted in \eqref{eq:ISRSGNmodel} through $S_k\left(f_1,f_2,f\right)=\sqrt[]{\frac{G_{k}(f_1)G_{k}(f_2)G_{k}(f_1+f_2-f)}{G_{k}(f)}}$ and $\tilde{L}_{k}=\sum_1^{k-1}L_k$. $L_k$ is the span length and $P_k$ the total input power of span $k$, $L_{\text{eff}}=\frac{1-e^{-\alpha \zeta}}{\alpha}$ and $\phi$ is a phase term resulting from dispersion \cite{SemrauSRS}.
\ 
\par 
For efficient integration, \eqref{eq:ISRSGNmodel} is fully implemented in hyperbolic coordinates yielding a computation time of 3-6~min on a 2.4~GHz single core CPU per evaluation $G\left(f\right)$.

%point to point-------------------------------------------------------%
\section{Point-to-point scenario}
The ISRS GN model is used to evaluate the NLI of the point-to-point link, from the BT network topology, Fig.~\ref{fig:networkscheme} (without ROADMs). For the point-to-point scenario, all channel slots are occupied and no channel is added or dropped at any ROADM. 
\ 
\par
To quantify the accuracy of \eqref{eq:ISRSGNmodel}, split-step simulations were performed for $\text{251}\times\text{40}$~GBd channels, occupying the entire C+L band (10.04~THz). A standard single mode fibre was considered with parameters $\alpha=0.2\frac{\text{dB}}{\text{km}}$, $D=17\frac{\text{ps}}{\text{nm}\cdot \text{km}}$, $S=0.067\frac{\text{ps}}{\text{nm}^2\cdot\text{km}}$, $\gamma=1.2\frac{\text{1}}{\text{W}\cdot\text{km}}$ and $C_r=0.028\frac{\text{1}}{\text{W}\cdot\text{THz}\cdot\text{km}}$. Gaussian modulation, uniform and Maxwell-Boltzmann shaped 64-QAM optimized for a SNR of 15~dB was used as transmitted symbols. The uniform channel launch power was 0~dBm which is the optimum launch power for the central channel in the presence of an Erbium-doped fibre amplifier with 5~dB noise figure.
\ 
\par
\begin{figure*}[]
\centering
\includegraphics[]{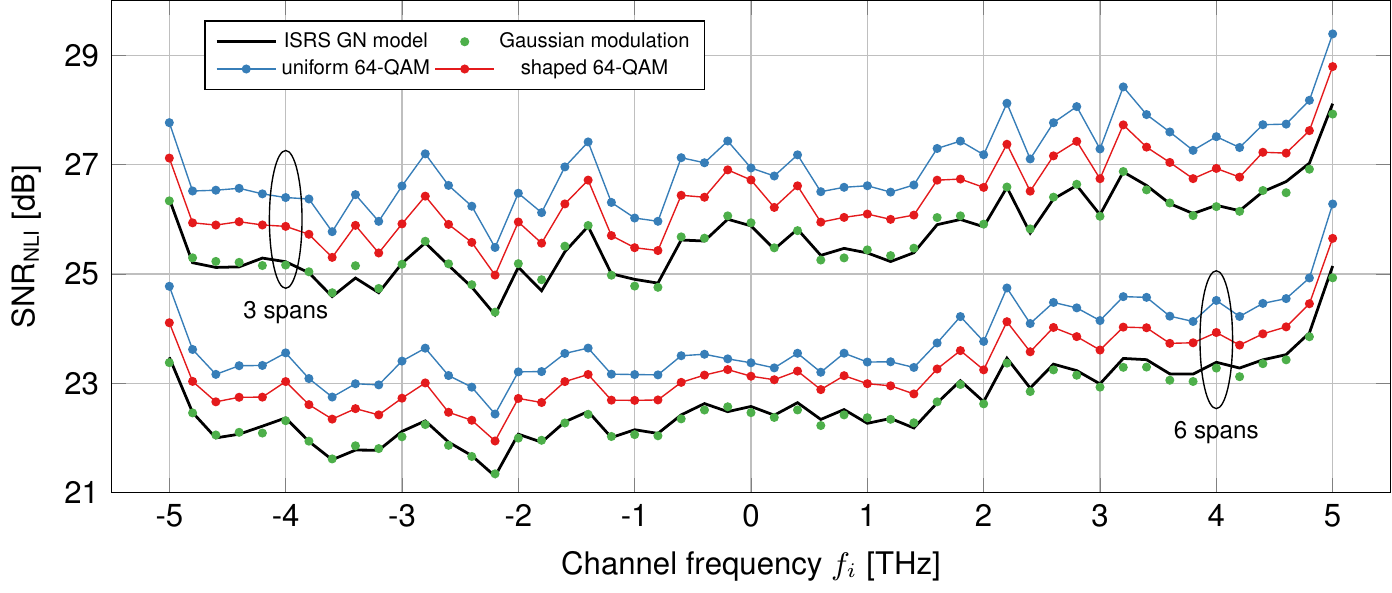}
\caption{\small Nonlinear performance of the signal channels considering a network scenario as illustrated in Fig. 1.}
\label{fig:network}
\end{figure*}

 
$2^{17}$ symbols were simulated in one transmission realization and four realizations were averaged to increase the simulation accuracy. The step size was logarithmically distributed and ISRS was implemented as a frequency dependent loss at every step. One realization took 42 hours (total simulation duration of 42 days for Fig. \ref{fig:straight} and 21 days for Fig.~\ref{fig:network}) on a state-of-the-art GPU.
\ 
\par 
The measured nonlinear SNR is shown in Fig.~\ref{fig:straight} without a) and with b) ISRS. The power difference after one span between the outer channels due to ISRS was 6.5~dB. The ISRS GN model shows excellent agreement in the case of Gaussian modulation with an average deviation of $<$0.1~dB. The average deviation to uniform and shaped 64-QAM is 1.9~dB and 1~dB after 3 spans and 1.6~dB and 0.8~dB after 6 spans. The tilt in Fig.~\ref{fig:straight}a) is a result of the dispersion slope $S$. Due to ISRS, the $\text{SNR}_{\text{NLI}}$ is changed by -2 to 1.8~dB as a consequence of the effective power amplification and depletion.
\ 
\par 
Eq. \eqref{eq:ISRSGNmodel} accurately predicts the impact of ISRS on the transmission performance in the point-to-point scenario.
%
%network------------------------------------------------------%
\section{Network scenario}
It is now assumed that channels are added and dropped at each ROADM to emulate a more realistic network environment as indicated in Fig.~\ref{fig:networkscheme}. Every fifth channel is kept along the entire path. Those 51 channels will be referred to as signal channels and their nonlinear performance will be evaluated. The remaining 200 channel slots act as interfering channels which are continuously dropped and added at each ROADM. 80\% of the interfering channels are dropped randomly (uniform) and channels are added randomly choosing an empty channel slot, until a 80\% network utilization is reached. In practice, unused channel slots are the result of variably loaded network spans.
\ 
\par 
The added channels exhibit a random power offset between $\pm1$~dB  with respect to the signal channels to account for non-ideal power equalization. The interfering channels use the same modulation format as the signal channels and were randomly pre-dispersed corresponding to a transmission distance between 0 and 1000~km to emulate the propagation from different lightpaths in the network. The wavelength dependent gain due to ISRS along a span was perfectly compensated to ease a comparison to the point-to-point case.
\
\par 
The nonlinear performance of the signal channels is shown in Fig.~\ref{fig:network}. The $\text{SNR}_{\text{NLI}}$ exhibits less average tilt than in the fully occupied case as less average optical power is transmitted. The average power difference between the outer channels due to ISRS was 5.2~dB. The fluctuation of the SNR (as high as 2~dB) is a consequence of the lightpath configuration. It can be seen that this fluctuation is smaller after 6 spans due to averaging. The average deviation between the ISRS GN model and Gaussian modulation is only 0.1 dB which can be considered negligible. The average deviation to uniform and shaped 64-QAM is 1.3~dB and 0.7~dB after 3 spans and 1.1~dB and 0.6~dB after 6 spans. The deviation is smaller than in the point-to-point case because the interfering channels exhibit in average more accumulated dispersion. More importantly, all formats exhibit similar SNR fluctuations as predicted by \eqref{eq:ISRSGNmodel}. 
\ 
\par 
We conclude that the ISRS GN model is accurate in the prediction of nonlinear performance in the presence of variable loaded network spans.

%------------------------------------------------- Section 8 -------------------------------------------------------%
\section{Conclusions}
A new model to accurately predict the impact of ISRS in fibre spans with variable wavelength channel loading has been proposed. The proposed model was compared to uniform and probabilistically shaped 64-QAM signals in C+L band (10~THz) point-to-point and network scenarios. In the network scenario, the average NLI deviation between model and simulation was only 1.1~dB for uniform and 0.6~dB for shaped 64-QAM, after a 742~km transmission. This makes the ISRS GN model an efficient tool for rapid performance estimation in ultra-wideband transmission networks.

%-------------------------------------------------- Section 9 -------------------------------------------------------%
\section{Acknowledgements}
% \begin{spacing}{1}%
\small{\textit{We thank Dr. D. Ives for useful comments. Support for this work is from UK EPSRC under DTG PhD studentship to D.~Semrau, INSIGHT project and UNLOC and TRANSNET Programme Grants.}}%
% \end{spacing}%
% \vspace*{-0.25cm}
%-------------------------------------------------- Literature -------------------------------------------------------%

\bibliographystyle{abbrv}
\begin{spacing}{1}

\end{spacing}
\vspace{-4mm}

%%%%%%%%%%%%%%%%%%%%%%%%%%%%%%%%%%%%%%%%%%%%%
%---------------------------------------------- End of Document -----------------------------------------------%
\end{document}